# A brief history of the origin of domesticated date palms

Gros-Balthazard, Muriel [1] & Flowers, Jonathan M.[1]

[1] Center for Genomics & Systems Biology, New York University Abu Dhabi, Abu Dhabi, United Arab Emirates

## Abstract

The study of the origins of crops is of interest from both a fundamental evolutionary understanding viewpoint, and from an applied agricultural technology perspective. The date palm (*Phoenix dactylifera* L.) is the iconic fruit crop of hot and arid regions of North Africa and the Middle East, producing sugar-rich fruits, known as dates. There are many different cultivars each with distinctive fruit traits, and there are many wild *Phoenix* species too, which in total form a complex of related species.

The understanding of plant domestication involves multiple disciplines, including phylogeography, population genetics and archaeology. In the past decade, they have prompted new discoveries on the evolutionary history of date palm, but a complete understanding of its origins remains to be elucidated, along with the genetic architecture of its domestication syndrome.

In this chapter, we review the current state of the art regarding the origins of the domesticated date palm. We first discuss whether date palms are domesticated, and highlight how they diverge from their wild *Phoenix* relatives. We then outline patterns in the population genetic and archaeobotanical data, and review different models for the origins of domesticated date palms by highlighting sources of evidence that are either consistent or inconsistent with each model. We then review the process of date palm domestication, and emphasize the human activities that have prompted its domestication. We particularly focus on the evolution of fruit traits.

# 1. Introduction

As a prelude to civilization, the advent of agriculture was a major advance in human history. The transition from hunter-gatherer societies to sedentary farming communities fostered innovations in plant cultivation and animal husbandry. From this Neolithic Revolution (Childe, 1936) emerged the earliest domesticated crops (see Box 1 for definitions), primarily cereals and pulses. While the earliest evidence of domestication of these crops dates to around 12,000 years ago, perennial fruit crops, including the major Mediterranean crops i.e. grapes, olives, and date palms, were domesticated later, from the end of the Neolithic period through the Bronze Age (6,000-3,000 BCE; Janick, 2005; Zohary and Spiegel-Roy, 1975). Today, these fruit crops are an essential part of the human diet and are critical for food security (Food and Agriculture Organization, 2013).

The history of the origin and diffusion of crops is at the forefront of genetic and archaeological research. Understanding where, when, and how plants were brought into cultivation sheds light on the history of our civilizations and the earliest stages of the domestication process. In the field of evolutionary biology, domestication provides a framework for answering broader evolutionary questions related to trait evolution and the origin of species (Ross-Ibarra et al., 2007; Zeder, 2017). In fact, Darwin (1859) developed his theory of evolution by natural selection to a large extent through observations of the effects of artificial selection in domesticated plants and animals. From an applied perspective, appraising diversity in wild and cultivated populations and characterizing the genetic basis of agronomic traits, such as fruit quality or disease resistance, provide key resources for plant breeding and crop improvement (Turner-Hissong et al., 2020).

The date palm (*Phoenix dactylifera* L.) is the major perennial fruit crop in hot and arid regions of the Old World and among the oldest cultivated fruit crops (Zohary and Spiegel-Roy, 1975). It belongs to the Arecaceae family and, along with 12 or 13 other interfertile species, comprises the Old World genus *Phoenix* (Barrow, 1998; Gros-Balthazard et al., 2020). This iconic species holds enormous economic, symbolic, and social importance throughout its traditional range of cultivation from Morocco in the west, across the Arabian Peninsula and to northwestern India in the east. Its sweet fruit, the date, has been consumed for millennia and served as a staple of subsistence farming and source of economic prosperity dating back to the earliest civilizations of the Middle East (Fuller and Stevens, 2019; Tengberg, 2012).

Date palms are dioecious, and phoeniciculture, the cultivation of date palm, includes a mix of clonal and sexual propagation. Traditionally, female cultivars are propagated asexually from basal offshoots to preserve fruit quality across generations. These cultivars are frequently given a name (e.g., 'Medjool' or 'Khalas'), and today there are more than 3,000 named varieties worldwide (Zaid and Arias-Jiménez, 1999) (Box 2). Small numbers of males are maintained in traditional palm groves to provide pollen that is used to manually pollinate the females. While this practice remains common today, the tremendous expansion of the date palm crop over the past 40 years was facilitated by propagation through tissue culture primarily via somatic embryogenesis (Zaid and Arias-Jiménez, 1999).

Date palms are the keystone species of the traditional system of oasis agriculture whose origin has been controversial for much of the last two centuries Archaeological remains indicate that domesticated date palms probably originated in the Middle East during the fourth millennium BCE, and later expanded throughout North Africa by the Roman period (Munier, 1973; Nixon, 1951; Tengberg, 2012). Many of the earliest archaeobotanical remains are concentrated in the Gulf Region the Tigris and Euphrates River valleys and early civilizations in this region documented a rich history of phoeniculture in ancient texts and iconography that date back to the end of the fourth millenium BCE (Tengberg, 2012).

Fresh insight into the domestication history of date palms has come from the fields of phylogenetics and population genomics where recent work has answered a number of fundamental questions (Gros-Balthazard et al., 2018). For instance, recent genomic studies suggest that relictual populations of the wild progenitor of date palms persists today in the Arabian Peninsula (Gros-Balthazard et al., 2017), support a role for introgressive hybridization in the diversification of the crop (Flowers et al., 2019) and provide evidence of extensive East-West exchanges of date palms at the height of the Roman Empire based on the genetic ancestry of date palms germinated from ancient seeds (Sallon et al., 2020). Despite these advances, many questions remain concerning the timing of events in the domestication history of date palms and how natural or artificial selection contributed to the domestication syndrome and trait evolution. Beyond the history of phoeniciculture, additional insight into date palm origins would provide insights into the foundation of oasis agrosystems and how humans colonized and adapted to the hot and arid regions of North Africa and the Middle East.

In this chapter, we aim to review the current knowledge about the origins of the domesticated date palms. We first discuss whether date palms are domesticated, and highlight their divergence to wild *Phoenix* relatives. We then outline patterns in the population genetic and archaeobotanical data, and review different models for the origins of domesticated date palms by highlighting sources of evidence that are either consistent or inconsistent with each model. We then review the process of date palm domestication, discuss the origins of human practices that promoted domestication, and highlight recent advances in studies of fruit trait evolution in this species.

## 2. Is the date palm a domesticated crop?

Domestication is an evolutionary process influenced by human activities whose outcome is animals and plants more suitable and profitable for human use (Harlan, 1992). The term domestication is sometimes confused with cultivation. The cultivation of a plant corresponds to its maintenance by humans. A plant can be cultivated without being domesticated; its cultivation does not necessarily lead to its domestication. On the other hand, Gepts (2004) considers that the cultivation stage is a prerequisite for domestication. However, by eliminating undesirable phenotypes directly in wild populations, domestication without cultivation (*in situ* domestication) can occur (Pickersgill, 2007).

There are several definitions of the term "domesticated plant". Harlan (1992) defines a fully domesticated plant as one that cannot survive without human intervention. Similarly, Purugganan & Fuller (2011) consider a plant to be domesticated when it depends on humans for reproduction and survival. This definition applies well to domesticated cereals and legumes that have characteristics that do not allow them to survive off the field (loss of seed dispersal in particular). However, most perennial plants, have evolved traits that distinguish them from their wild ancestors (the "domestication syndrome"; Box 1), and yet can survive and reproduce without human intervention. Therefore, this definition is overly restrictive in the context of perennial crops as most would not be considered domesticated. In anthropology, it is human's perception of wild and cultivated plants that legitimizes the use or not of the term "domesticated plant".

Here, we adopt the definition of domesticated plants proposed by Meyer et al. (2012): " 'domesticated' refers more generally to plants that are morphologically and genetically distinct from their wild ancestors as a result of artificial selection, or are no longer known to occur outside of cultivation". By this definition, date palms are considered domesticated, though some local communities consider them to be wild in their gardens (e.g. Tuaregs of the Tassili n'Ajjer, Algeria; Battesti, 2004).

## 3. *Phoenix* wild relatives and the wild progenitor of date palms

Crop wild relatives consist of those phylogenetically related taxa that share a recent common ancestor with the domesticated species. These species are of tremendous interest both in evolutionary studies of domestication (see below) and in applied contexts owing to opportunities for crop improvement. For example, many crop wild relatives are more tolerant of biotic and abiotic stresses and thus represent a reservoir of diversity for breeding and crop improvement (Burgarella et al., 2019; Migicovsky and Myles, 2017; Zhang et al., 2017). In date palms, vanishingly little is known about stress tolerances and other traits of interest in 13 wild relatives of *P. dactylifera* and most of what is known is based on observations of habitat occupancy such as the preference of *P. theophrasti* for coastal areas and ability to survive salt water exposure (Barrow 1998). Here we restrict our discussion to ways in which studies of *Phoenix* wild relatives have informed the domestication of date palm, but there is a great need for experimental assessment of trait diversity in these species for advances both in applied and evolutionary studies (Hazzouri et al. 2020).

## 3.1 Hybridization in the *Phoenix* genus

Species in the *Phoenix* genus have weak barriers to gene flow and frequently hybridize in anthropogenic contexts (reviewed in Gros-Balthazard, 2013). This has fueled speculation that domesticated date palms may have a hybrid origin (reviewed in Pintaud et al., 2010) or that hybridization may have been a source of genetic variation during the course of domestication (Gros-Balthazard et al., 2017; Hazzouri et al., 2015; Pintaud et al., 2010). As discussed below, the hypothesis that the ancestral gene pool of domesticated date palms is the fusion of two *Phoenix* species is not supported by existing data, although there is growing support for the hypothesis that regional population(s) of *P. dactylifera* may have been subject to introgressive hybridization (see below).

## 3.2 "Wild" versus feral date palms

The adjective "wild" describes either the ancestor (i.e., progenitor) of a domesticated crop or a phylogenetic relative of the domesticated species. Unfortunately, although the term "wild" has clear meaning in this context, the term is often incorrectly applied to uncultivated populations. Indeed, in date palms, like many perennials, untended stands become established either when a farm is abandoned, or when individuals 'escape' cultivation via germination from seed. Uncultivated stands established in these ways are not genuinely wild, but are feral (Box 1) and may be mis-characterized as a relictual population of the ancestor of domesticated date palms due to shared phenotypic traits between this wild species and feral forms. Although such untended groups of date palms can be found throughout North Africa and Western Asia (Gros-Balthazard et al., 2018: Fig. 2), whether they are feral or wild is unknown in most cases.

These populations may be critical to understanding key aspects of the origin and spread of date palms, as shown by the recent study of feral populations scattered in the desert near Siwa oasis, Egypt (Gros-Balthazard et al., 2020). Unfortunately, most studies have focused on the cultivated germplasm and these populations of uncultivated populations remain poorly understood as a result.

## 3.3 The wild progenitor

The wild progenitor of domesticated date palms has long been a mystery in the study of the crop, and until recently has remained unknown. The issue of distinguishing wild and feral individuals has been at the forefront of this debate, and prior to the application of genetic data, it was unknown whether date palm origins trace to a wild population of *P. dactylifera,* an extant wild relative such as *P. sylvestris*, , or hybridization between two or more *Phoenix* species (Pintaud et al., 2010; Zaid and Arias-Jiménez, 1999).

Independent studies have now excluded the possibility that the domesticated population of *P. dactylifera* was the product of hybridization between two wild *Phoenix* species and established the identity of the progenitor. A genetic study of the *Phoenix* genus established that *P. dactylifera* is sufficiently diverged from its congenerics that none of these species were the wild progenitor and it was instead likely that date palms were domesticated from a wild population of the same species (Pintaud et al., 2010). A subsequent study supported this hypothesis. Gros-Balthazard et al. (2017) presented evidence that uncultivated stands of *P. dactylifera* growing in isolated areas of the Hajar Mountains of Oman are wild, not untended populations of feral palms. This provided the strongest support yet that a relictual population of the wild progenitor of domesticated date palms persists to the present-day and provides new opportunities to study the ancestor of the crop.

Description of the native range of a wild progenitor species is a major issue when studying crop origins because domestication centers typically falls within or at the edge of the historic distribution. Having solely been described from the Hajar Mountains of Oman, the native range of the wild progenitor of date palms is unknown. Many uncultivated populations could also represent wild populations and warrant verification through genetic and morphologic analyses (Gros-Balthazard et al., 2018, 2017).

Ancient evidence for wild date palms can be found as archaeobotanical remains in Western Asia that date to the Holocene. Both > 50 ka pollen and ca. 46 ka phytoliths attributed to the date palm were found in the Shanidar cave in Iraq (Henry et al., 2011; Solecki and Leroi-Gourhan, 1961). In Iraq, date palm phytoliths were retrieved from sediments dated to ca. 10,000 BCE (Altaweel et al., 2019). Excavations in the Levantine regions yielded a ~19,000 year old burnt stem and ~49-69 ka phytoliths (Henry et al. 2004; Liphschitz and Nadel 1997) that were attributed to *Phoenix dactylifera*, based on it being the only *Phoenix* growing in the region today,. However, these finds cannot formally be identified to the species level given a lack of diagnostic criteria to distinguish species (Gros-Balthazard et al., 2020). Palm phytoliths have also been found in Jebel Faya, United Arab Emirates, in sediments dated to ca. 125 kya, although it is unclear whether they belong to date palm or *Nannorrhops ritchiana* (Bretzke et al., 2013). In the southwest of the Kingdom of Saudi Arabia, palm phytoliths were discovered in c. 80 kya deposits; these may be attributable to the date palm but the rarity in the assemblage may suggest long distance transport by wind (Groucutt et al., 2015). In Africa, pre-Neolithic remains of *Phoenix* are almost nonexistent. The sole evidence is from the Egyptian oasis of Kharga, where carbonized seeds and a fossilized leaf of *Phoenix* were recovered from Pleistocene deposits (Caton-Thompson and Gardner, 1932; Gardner, 1935); nevertheless, only the sediment has been dated and the age of the *Phoenix* remains is unknown and they could represent more recent contaminating remains.

Based on these considerations, it seems likely that wild *Phoenix dactylifera* are native to Western Asia, although the precise distribution is unknown and a larger historical distribution, covering all or parts of North Africa cannot be ruled out.

Prior to cultivation and domestication, wild date palms had been exploited for millennia. For example, it was part of Neanderthals' diet 50,000 years ago, as evidenced by phytoliths found in dental calculus from teeth recovered in Shanidar cave, Iraq (Henry et al., 2011). Earliest evidence of date palm exploitation by modern humans have been found on two sites on the Gulf coast: Dalma Island, United Arab Emirates (Beech and Shepherd, 2001) and Sabiyah, Kuwait (Parker, 2010) and date back to approximately ~5,000 BCE (Fig. 1).

# 4. Origins and diffusion of the date palm

The origins of domesticated date palms have been controversial since the beginning of the nineteenth century (Tengberg, 2003). Scholars have advanced different theories about the wild progenitor species and debated the location of center(s) of origin, the number of domestication events, and how hybridization may have contributed to the origin and diversification of the crop. Here we briefly summarize some of the most prominent ideas and discuss in more detail hypotheses that are supported by existing data.

## 4.1. Origin hypotheses

### Geographic origins

The diversity of hypotheses that have been advanced concerning the geographic origin of date palms reflect the poor understanding of both the identity and historical range of the ancestral species. In fact, origin hypotheses have been proposed for most regions where date palms are cultivated in the Old World. For example, authors have proposed geographic origin scenarios ranging from Northwest Africa in the West, to Ethiopia in the South, to the Western India in the East, and many locales in between (reviewed in Barrow, 1998; Goor, 1967; Munier, 1981; Tengberg, 2003). Most recent scholars, however, have considered regions of the Middle East to be the most likely origin, in keeping with available evidence on the prehistoric distribution of *P. dactylifera*, ancient written records, archaeological remains and archaeobotanical finds (Fuller and Stevens, 2019; Tengberg, 2012, 2003; Zohary and Spiegel-Roy, 1975).

## Number of domestication events

Whether date palms were independently domesticated one or more times is a source of ongoing debate as it is in the studies of many crop species (i.e. in olives, Besnard et al., 2018). These controversies emerge in the field of population genomics from challenges associated with reconstructing complex historical events from patterns of genetic ancestry and population structure and associating these patterns with human-mediated selection and other domestication-related activities. This problem is exacerbated in relatively poorly studied crops, including date palms.

Population genetic studies of regional populations of domesticated crops often find distinct ancestries that are commonly interpreted as evidence of multiple domestication events. Choi et al. (2017) argued, however, that evidence of distinct genetic ancestries may not reflect independent *de novo* domestication, but may instead represent single domestication with multiple origins. Multiple origins in this context refers to the number of independent ancestral gene pools from which a crop is derived irrespective of whether the ancestral populations were independently domesticated.

Identifying multiple origins is relatively straightforward with population genomic data, but it is considerably more difficult to determine if different sources of genetic variation were subject to independent selection regimes that are a hallmark of multiple domestication. In the case of rice, unidirectional gene flow of functional alleles at key loci that control domestication-related traits provide evidence of a single origin of domestication traits. This supports a single domestication despite multiple origins apparent in the distinct genetic ancestries of subspecies of domesticated rice (Choi et al., 2017). Thus, genomic data frequently support multiple origins, but evidence of multiple domestications remains tenuous in most study systems.

Cultivated date palms are geographically structured into eastern and western populations (Arabnezhad et al., 2012; Hazzouri et al., 2015; Mathew et al., 2015; Zehdi-Azouzi et al., 2015), with additional minor divisions within each of these regions (Gros-Balthazard et al., 2020; Mohamoud et al., 2019; Zango et al., 2017; Zehdi-Azouzi et al., 2015). Population genetic studies have in some cases interpreted this geographic structure as evidence of multiple domestications ranging from two to four events. For example, Zehdi-Azouzi et al. (2015) and Mathew et al. (2015) proposed that the geographic structure detected between North African and Middle Eastern date palms supports two domestications. However, the geographic structure observed in date palms may also be explained by a multiple origin, single domestication model. For example, it is possible that domestication-related traits were selected in the East and alleles controlling these traits were later introduced to a proto-domesticated or wild population in North Africa. This scenario could account for population structure without multiple *de novo* domestications.

Determining the number of domestication events will continue to be challenging in date palms. There is a need to continue to evaluate evidence that date palms are the product of multiple domestications or if factors such as hybridization could be the source of the distinct genomic ancestries in geographic populations of date palm (see below) without the need to invoke multiple domestications. There is also a need to consider that domestication of date palms may not be attributable to origins from well-defined cultivation centers, but is a geographically diffuse process as has been suggested for other fruit crops (Miller and Gross, 2011).

## 4.2. Introgressive Hybridization

The North African population of date palms is of particular interest in the context of introgressive hybridization. This population is genetically distinct from the Middle East (Arabnezhad et al., 2012), consists of multiple sub-populations (Gros-Balthazard et al., 2020; Zango et al., 2017; Zehdi-Azouzi et al., 2015), has at least 20% higher genetic diversity than populations in the Arabian Peninsula and elsewhere in the Middle East (Gros-Balthazard et al., 2017; Hazzouri et al., 2015), and has distinct and deeply divergent chloro- (i.e. Pintaud et al., 2013) and mito-types (Flowers et al., 2019; Mohamoud et al., 2019) that are found at high frequency and largely restricted to North Africa. Hazzouri et al. (2015) speculated that these patterns may be at least partially explained by introgressive hybridization with a wild relative.

Evidence supporting introgression from wild *Phoenix* in North Africa has begun to accumulate (Flowers et al., 2019; Gros-Balthazard et al., 2017; Gros-Balthazard et al., 2020). Direct evidence of introgression was first reported by Flowers et al. (2019) using whole genome resequencing of date palms and its closest wild relatives. In this study, explicit tests of admixture (e.g., ABBA-BABA and $f_3$ tests) supported introgression between the North African population and a closely related congeneric *P. theophrasti*, or a *P. theophrasti*-like population (i.e., a possibly extinct species or population that is closely related to *P. theophrasti* and may have the direct source of introgressed alleles). Introgression was further supported by the segregation of *theophrasti* alleles in North Africa, population modelling that included admixture between North African date palms and *P. theophrasti*, haplotype sharing in introgressed genomic regions, and patterns of linkage disequilibrium that are consistent with a recent history of admixture with a distant population. In addition, genomewide estimates of population divergence supported reduced divergence in the North Africa-*theophrasti* comparison versus Middle East-*theophrasti*.

Evidence of introgressive hybridization with a *P. theophrasti*-like population has also been implicated in a microsatellite-based study of date palms in the Siwa Oasis, Egypt. Gros-Balthazard et al. (2020) surveyed more than a hundred cultivated and feral date palms, and found that both shared alleles with *P. theophrasti* to a greater extent than samples from the Middle East. Interestingly, the degree of allele sharing was higher in the samples from Siwa oasis, Egypt, than in the other samples from North Africa, and especially high in the feral accessions. Whether this intriguing pattern reflects patterns of selection favoring "wild" alleles in feral populations or reflects differences in demographic history and the history of hybridization is an area of ongoing research.

How much of the North African genomic ancestry traces to *P. theophrasti* or the *theophrasti*-like population? Estimates of *P. theophrasti* ancestry in North African date palms was estimated at 5-18% (Flowers et al., 2019). The upper bound of this estimate is remarkably consistent with an estimate of ancestry from an unsampled "ghost" population (18%) in North African date palms (Gros-Balthazard et al., 2017). The signature of introgression is strongest in cultivars from the Maghreb that have 15-18% of their genomic ancestry that traces to *P. theophrasti*, while cultivars from Egypt and Sudan showed a smaller (5%) ancestry fraction.

*P. theophrasti* is a tertiary relict with present-day distribution limited to Crete and the Aegean Sea region (Boydak, 2019; Vardareli et al., 2019), but once may have had a broader geographic range in the Eastern Mediterranean (Fuller and Stevens, 2019; Kislev et al., 2004). The fact that the current range of *P. theophrasti* does not overlap with the range of date palm remains a source of uncertainty concerning the geography of introgressive hybridization between these species (Flowers et al., 2019).

Questions related to the geographic context of hybridization, the age of the introgression event(s), whether hybridization was human-mediated or the product of natural events are the subject of continuing work. These questions are currently being addressed through expanded surveys of genomic variation, studies of ancient DNA, and population genetic modelling. Other areas of active investigation include studies of other *Phoenix* wild relatives to determine the extent to which hybridization with additional species may have contributed to the diversity of date palms (Pintaud et al., 2010).

## 4.3. Models of domesticated date palm origins

We have outlined patterns in the population genetic, archaeobotanical and other sources of data that are consistent with some models for the origins of domesticated date palms and inconsistent with others. Here we expand on geographic models of date palm domestication and highlight sources of evidence that are either consistent or inconsistent with each model. We outline the simplest models that could account for current data.

## Expansion model

Prior to the availability of genetic data, the preferred model for date palm origins proposed a domestication center in the Middle East followed by range expansion (Tengberg, 2003). Under this model, the current range of cultivation in the old world--from North Africa to Pakistan--resulted from westward dispersal from a domestication center in the East via trade routes traversing the Sahara (Munier, 1973) or sea-faring routes across the Mediterranean (Nixon, 1951). Under this model, any population of wild *Phoenix* that may have occupied habitats in North Africa was replaced without admixture.

Key elements of this model are supported by archaeobotanical and archaeological remains. For example, various sources support a thriving date palm culture in ancient Mesopotamia and the Upper Gulf Region by the early Bronze Age (late 4th/early 3rd millennia B.C.) (Tengberg, 2012)(Fig. 1). By contrast, there is limited evidence of ancient date palm cultivation or wild *Phoenix* remains in Northern Africa. The earliest evidence of cultivation dates to the New Kingdom, mid-second millennium BCE, in the Nile River Valley (Popenoe, 1924; Tengberg and Newton, 2016), and from the beginning of the 1st millennium BCE in Fezzan in modern-day Libya (Pelling, 2005)(Fig. 1). Prior to that, there are only sparse remains in Egypt (reviewed in Gros-Balthazard et al., 2020; e.g., Giza, 2,700-2,100 BCE; Malleson, 2016; Malleson and Miracle, 2018) and no reliable remains recorded further west (Flowers et al., 2019). The earliest remains in the Maghreb date to much later, first appearing at the Roman settlement of Volubilis (Morocco) in the Classical Period (ca. 400-100 BCE; Flowers et al., 2019; Fuller and Pelling, 2018)(Fig. 1). The expansion model is the simplest way to account for the absence of wild *Phoenix* remains in North Africa and the disparity in ages of remains in North Africa and the Upper Gulf/Eastern fringe of the Fertile Crescent (Tengberg, 2012).

The expansion model is attractive in its simplicity, but population genetic studies indicate a more complex history. Arguably, the two most difficult patterns to reconcile with a simple expansion is the higher nucleotide diversity in the Western population and evidence of introgression in North Africa. The difference in diversity is inconsistent with a population genetic bottleneck that presumably would have accompanied the founding of a new population on the African continent (Hazzouri et al., 2015), while introgression is inconsistent with a simple expansion. Despite these observations, genetic data provide indirect support for gene flow from the Middle East to North Africa consistent with a westward expansion. For example, asymmetrical gene flow into North Africa has been proposed to account for the low to moderate frequency of the "Eastern" cpDNA and Y-chromosome haplotypes in North Africa, but the near absence of the "Western" cpDNA and Y-chromosome haplotypes in the Arabian Peninsula and elsewhere in the Middle East (Cherif et al., 2013; Hazzouri et al., 2015; Mathew et al., 2015; Zehdi-Azouzi et al., 2015). Second, Gros-Balthazard et al. (2017) and Flowers et al. (2019) estimated that ~82% of North African ancestry can be traced to the Middle Eastern date palm. While there are alternative explanations for these patterns, the simplest explanation is gene flow of Eastern alleles into North Africa at a time of geographic expansion of the crop.

## "Leaky" expansion model

The "Leaky" expansion model includes the westward movement of Middle Eastern date palm into North Africa as proposed by the expansion model, but also proposes admixture with a *Phoenix* wild relative such as *P. theophrasti* or a *theophrasti*-like population. Although two independent reports support hybridization between date palm and *P. theophrasti* (Flowers et al., 2019; Gros-Balthazard et al., 2020), details of the nature of hybridization are unclear and it is presently difficult to distinguish among various competing scenarios (Fig. 1). The disjunct present-day distributions of cultivated date palms and *P. theophrasti* add uncertainty to the geography of hybridization in the past.

For example, one possibility is that introgression occurred prior to range expansion of the Middle Eastern crop outside of Africa, where the two species ranges may have once come into contact (Fig. 1A; Flowers et al., 2019). *P. dactylifera* X *P. theophrasti* hybrid populations on Crete (e.g., Almyros; Flowers et al., 2019) and possibly Turkey (Boydak and Barrow, 1995) represent extant examples of such admixed populations.

An alternative model incorporates the expansion of Middle Eastern date palms to North Africa, but proposes that the introgressive hybridization occurred *in situ* with a resident *theophrasti* or *theophrasti*-like population (Fig. 1B). This model could explain *theophrasti* introgression signatures in North Africa. However, the North African populations is segregating a high frequency cpDNA and mtDNA haplotypes that are not shared with *theophrasti* (hence the reference to "theophrasti-like") and found at very low frequency elsewhere (< 4%; Zehdi-Azouzi et al., 2015). This "North African" haplotype may elude to a more complex history such as an additional source of ancestry in this population (Fig. 1C). An alternative hypothesis is that this haplotype is not North African *per se*, but traces in origin to the Arabian Peninsula (where it is found at low frequency) and its current frequency on the African continent is the product of demographic events following range expansion. Additional population genetic modelling is required to distinguish among these possibilities.

Both models illustrated in Fig. 1B and Fig. 1C propose a resident wild population(s) of date palm in North Africa of which there is little or no support in the archaeobotanical record. However, there is a 20th century description of *P. atlantica* var. Moroccana in Morocco (Chevalier, 1952) and more recent unverified reports of possible wild date palms in isolated areas (Zehdi-Azouzi et al., 2016; Zohary et al., 2012). Whether these reports describe feral date palms or wild *Phoenix* is unknown, but it remains possible that the archaeobotanical record is incomplete and that such a "ghost" population could be a source of genetic ancestry in North African cultivars (Gros-Balthazard et al., 2020).

Finally, Cyrenaica (modern-day Libya) and Crete had close geo-political ties and comprised a senatorial province during the Roman empire (Chevrollier, 2016). It is conceivable that ancient trade routes such as those connecting Crete to continental Africa may have facilitated the transport of *P. theophrasti* to North Africa (e.g. for construction materials or other uses). This is one example of related hypotheses that posit both the expansion of date palm and transport of *P. theophrasti* to the African continent where hybridization may have occurred (Fig. 1D).

## Additional complexity

There are features of the population genomic data that are not easily explained by the above models. For example, Flowers et al. (2019) broke the genome into segments and estimated the introgression fraction between *P. theophrasti* and North African date palm in each genomic region. Segments of the genome with higher introgression fraction had correspondingly higher nucleotide diversity in North Africa consistent with introgression from *P. theophrasti* contributing to the elevated nucleotide diversity in this population (Gros-Balthazard et al., 2017; Hazzouri et al., 2015). Other regions in the genome did not show evidence of introgression between North Africa and *P. theophrasti*. The potentially revealing observation is that these non-introgressed genomic segments still showed higher nucleotide diversity on average in North Africa compared to the Middle East, albeit not as pronounced the difference observed in genomic regions introgressed by *P. theophrasti* alleles. This suggests that *P. theophrasti* introgression may not fully account for the higher diversity in North Africa.

What does this suggest about models of population history? The above observation is inconsistent with a simple "leaky" expansion model because there is no evidence of a population bottleneck even in non-introgressed genomic regions. To account for this pattern, it may be necessary to consider factors that increase North African diversity relative to the Middle East in addition to introgression from *P. theophrasti*. For example, additional sources of genetic ancestry in North Africa or bottlenecks, inbreeding, or stronger diversity-reducing effects of natural or artificial selection in the Middle East could account for this pattern (Flowers et al., 2019). These and other population genetic scenarios would benefit greatly from extensive population genetic modelling of the domestication history of date palms.

# 5. The process of domestication

The domesticated date palm finds its origins in the human practices that have shaped the genetic makeup of *Phoenix dactylifera* and favored desirable phenotypes through time, leading to a distinction between wild and domesticated populations (Box 3).

The initial cultivation of date palms by early farmers represents the earliest stage of domestication. The practices adopted by these pioneering farmers were primarily related to reproduction and propagation (detailed below), but also include practices such as leaf removal, pruning of fruit bunches, and selection for desirable traits such as larger and better tasting fruit. Here we detail what is known about these processes in the context of date palm domestication and discuss recent advances in studies of the evolution of date palm fruit traits (section 5.2).

## 5.1. The origins of cultivation practices

### Vegetative propagation

One of the most important events in the domestication of fruit-bearing crops was the capacity to vegetatively propagate the crop. In perennial plants, 75% of cultivated species are propagated clonally (by grafting, cutting, layering or planting offshoots), which confers several advantages over sexual reproduction (McKey et al., 2010; Miller and Gross, 2011).

Date palms are vegetatively propagated using offshoots growing at the base of the palm, that are planted independently of the mother palm and themselves provide offshoots. Vegetative propagation holds several advantages over sexual reproduction. First, it allows ascertaining that a female will grow while when planting seed, the ratio male/female is 50/50. Second, the juvenile phase is reduced and date palms will provide more rapidly the seeked fruits. Lastly, and more importantly, vegetative propagation allows the selection of individuals with interesting phenotypic traits (especially the fruits of the desired quality) and their identical reproduction. Only in rare cases does sexual reproduction yield date palms that are as high quality as or more interesting than the parents (4‰ according to Peyron, 2000).

Clonal propagation of date palms is a very ancient practice. The oldest evidence of it is from a soft-stone vessel that probably shows a scene of date palm propagation by offshoot dating back to the second half of the third millennium BCE (Tengberg, 2012) (Fig. 1).

### Manual pollination

The second main farming practice associated with phoeniciculture is manual pollination (Roué et al., 2015). Pollination of the female inflorescence is necessary for fruit development and ripening, as already known by the Sumerians in Iraq, ca. 2300 BCE (Janick, 2005). In the wild, it is taken care of by either insects or wind (Barrow, 1998). But for the oasis system to be efficient, despite the scarcity of water, irrigable lands, and manure, oasis communities maintain 95%–99% of female palms. Opting for such an artificial sex ratio requires hand pollination as the very low proportion of males in the gardens do not allow the natural pollination of all female flowers, leading to a loss in fruit yield.

Manual pollination was probably already in use in southern Mesopotamia from the late 4th millennium BCE (Landsberger, 1967; Tengberg, 2003). It's first mention in texts got back to the 18th century BCE, in the famous code of Hammurabi, a Babylonian king, where it was associated with religious practices and later this practice was prominent in Assyrian iconography (Sarton, 1934)(Fig 1).

### Other cultivation practices

The domestication of the date palm is associated with the origins of oasis agrosystems, of which it is the keystone species (Tengberg, 2012). A major constraint in this environment is the scarcity of water resources and the high intra- and inter-annual variability in their availability. The development of oasis agrosystems has thus been accompanied by the development of irrigation systems, such as the *falaj* or *qanat* system. These systems are hardly precisely dated, and where they originate is unknown. Nevertheless, they occur since at least the Iron Age in Southeast Arabia and could be as old as the Bronze Age (reviewed in Charbonnier, 2013).

## 5.2. The evolution of date palm fruit traits

Fruit quality traits such as size, shape, color, flavor, and texture are primary targets of selection in fruit crops and a key component of the domestication syndrome in date palm (Box 1). Clues to the origin and evolution of date palm fruit traits come from studies of the diversity of date palm fruits (Zaid and Arias-Jiménez, 1999), comparisons of fruits of domesticated *P. dactylifera* and its wild relatives (Amorós et al., 2009), changes in seed size and shape apparent in the archaeobotanical record (Figure 3; Fuller, 2018; Terral et al., 2012), and genetic studies of fruit traits (Hazzouri et al., 2019, 2015).

A visit to the souks of the Arabian Peninsula or North Africa and the diversity in color, shape, size, texture and taste of date palm fruits is on full display. At the fresh ('khalal') stage when select varieties are consumed, the fruits range from pale yellow (e.g., 'Barhee') to deep red (e.g., 'Hayany'), with many intermediate colors including shades of pink and orange. Even after ripening is complete, color remains an important distinguishing feature with exceptional varieties such as 'Ajwa' assuming a dark black color at the dry stage.

Many other varieties are recognizable based on their shape, size and flavor. Shapes include round ('Braim'), ovoid ('Khalas'), and elongate ('Deglet Noor') and sizes range from small fruited varieties (e.g., 'Lulu') to large fruited varieties (e.g., 'Anbar' and 'Medjool'). Each variety is also characterized by distinctive texture and taste. Many North African varieties are dry, while those of the Arabian Peninsula and the Middle East are more typically semi-dry or soft. These textures are correlated with sugar composition with dry varieties frequently retaining sucrose through ripening whereas semi-dry and soft varieties typically hydrolyze sucrose to glucose and fructose (Dowson and Aten, 1962). An exceptional example of a sucrose-type variety is 'Sukary', which deposits sucrose in high concentrations in the fruit pulp and is valued in Arabia and beyond for its unusually sweet taste.

Understanding the origins and selective mechanisms acting on fruit traits has great potential to inform understanding of date palm domestication. As a first step, characterization of the genes and mutations that control these traits are fundamental to understanding the process of date palm domestication. For example, Hazzouri et al. (2015) reported a statistical association between genotypes at the *VIRESCENS* locus in date palm and 'khalal' stage fruit color. *VIRESCENS* codes for an R2R3-MYB transcription factor that is expressed in the fruit and activates anthocyanin biosynthesis. A genomewide association study (GWAS) subsequently confirmed that *VIRESCENS* is the primary locus that controls color variation in date palm fruits (Hazzouri et al., 2019).

Candidate mutations for the causal polymorphisms for fruit color have been identified at the *VIRESCENS* locus. Hazzouri et al. (2015, 2019) reported a polymorphic retrotransposon insertion (named *Ibn Majid* after the 15th Century Arab navigator) in the third exon that disrupts the open reading frame. This mutation acts as a dominant negative that suppresses anthocyanin production. Hazzouri et al. (2019) reported a translation initiation codon mutation (ATG->ATA) that acts a recessive loss-of-function mutation. These two mutations in *VIRESCENS* support a model that accounts for much of the variation in fruit color. The relatively uniform yellow fruit color in wild stands of the closest relatives of date palm (e.g., *P. sylvestris* and *P. theophrasti*), and multiple independent mutations within date palm suggest that fruit color was selected during date palm domestication.

Sugar composition is a prominent trait that varies among date palm cultivars. Date palms deposit large concentrations of sugar during fruit development with as much as 80-85% in the form of sucrose at the fresh, or 'khalal', stage (Chao and Krueger, 2007). At the onset of ripening, many varieties such as 'Khalas' invert sucrose to reducing sugars to the extent that there is little or no sucrose retained in the tamar (dry) stage. Other varieties such as 'Sukkari' retain sucrose in high concentrations in the dry stage. Hazzouri et al. (2019) reported a quantitative trait locus (QTL) on linkage group 14 that controls variation in this trait in date palm fruits. They found that within the QTL is a cluster of cell wall invertases and an alkaline/neutral invertase and reported what appear to be multiple deletions in this region including the homozygous deletion of a cell wall invertase in many of the sucrose-type varieties. This suggests that sucrose-rich varieties may have evolved via selection on loss-of-function alleles at the invertase locus. A similar finding was reported in a subsequent independent study (Malek et al., 2020).

Amoros et al. (2014) surveyed sugar composition and other compounds in developing fruits of date palms and their *Phoenix* wild relatives. *P. dactylifera* were the only *Phoenix* species surveyed to deposit large amounts of sucrose in 'khalal' stage fruit, whereas only trace amounts were reported for *P. loureiroi*, *P. canariensis*, *P. roebelenii*, and *P. reclinata*. This suggests that the process of sucrose deposition in date palms is unique to domesticated date palm and further suggests a two phase model for the origin of the sweet sucrose-rich varieties. In the first stage, date palms were selected by early farmers to increase the deposition of sucrose in early stages of fruit development perhaps to offset the acidic taste at the fresh stage. In the second stage, at the 'sukary'-type varieties were selected to retain sucrose throughout the ripening process. Studies such as those of Amoros et al. (2014) that include wild relatives of date palm are rare but important for studies of domestication.

Future research on the fruit color and sugar QTL regions will focus on the origins of the alleles and the selective forces (e.g., soft or hard sweeps) that may be operating at these loci.

# 6. References

Alcántara, J.M., Rey, P.J., Alcantara, J.M., Rey, P.J., 2003. Conflicting selection pressures on seed size: Evolutionary ecology of fruit size in a bird-dispersed tree, Olea europaea. J. Evol. Biol. 16, 1168–1176. https://doi.org/10.1046/j.1420-9101.2003.00618.x

Altaweel, M., Marsh, A., Jotheri, J., Hritz, C., Fleitmann, D., Rost, S., Lintner, S.F., Gibson, M., Bosomworth, M., Jacobson, M., Garzanti, E., Limonta, M., Radeff, G., 2019. New Insights on the Role of Environmental Dynamics Shaping Southern Mesopotamia: From the Pre-Ubaid To the Early Islamic Period. Iraq 1–24. https://doi.org/10.1017/irq.2019.2

Amorós, A., Pretel, M.T., Almansa, M.S., Botella, M.A., Zapata, P.J., Serrano, M., 2009. Antioxidant and Nutritional Properties of Date Fruit from Elche Grove as Affected by Maturation and Phenotypic Variability of Date Palm. Food Sci. Technol. Int. 15, 65–72. https://doi.org/10.1177/1082013208102758

Arabnezhad, H., Bahar, M., Mohammadi, H.R., Latifian, M., 2012. Development, characterization and use of microsatellite markers for germplasm analysis in date palm (*Phoenix dactylifera* L.). Sci. Hortic. (Amsterdam). 134, 150–156. https://doi.org/10.1016/j.scienta.2011.11.032

Aruz, J., 2003. Art of the First Cities: The Third Millennium B.C. from the Mediterranean to the Indus (Catalogue of an exhibition held at the Metropolitan Museum of Art from May 8 to Aug. 17, 2003). Metropolitan museum of art, New York and Yale University Press,

New Haven.

Balick, M.J., 1984. Ethnobotany of palms in the Neotropics. Adv. Econ. Bot. 1, 9–23.

Barrow, S., 1998. A monograph of *Phoenix* L. (Palmae: Coryphoideae). Kew Bull. 53, 513–575.

Battesti, V., 2004. Odeur sui generis, Le subterfuge dans la domestication du palmier dattier (Tassili n'Ajjer, Algérie), in: Bonte, P., Brisebarre, A.-M., Helmer, D., Sidi Maamar, H. (Eds.), Anthropozoologica — Domestications animales : dimensions sociales et symboliques (Hommage à Jacques Cauvin). Publications Scientifiques du Muséum, Paris, pp. 301–309.

Beech, M., Shepherd, E., 2001. Archaeobotanical evidence for early date consumption on Dalma Island, United Arab Emirates. Antiquity 75, 83–89.

Besnard, G., Terral, J.-F., Cornille, A., 2018. On the origins and domestication of the olive: a review and perspectives. Ann. Bot. 121, 385–403. https://doi.org/10.1093/aob/mcx145

Bolmgren, K., Eriksson, O., 2010. Seed mass and the evolution of fleshy fruits in angiosperms. Oikos 119, 707–718. https://doi.org/10.1111/j.1600-0706.2009.17944.x

Boydak, M., 2019. A new subspecies of Phoenix theophrasti Greuter (*Phoenix theophrasti* Greuter subsp. *golkoyana* Boydak) from Turkey. Forestist 69, 133–144. https://doi.org/10.26650/forestist.2019.19016

Boydak, M., Barrow, S., 1995. A new locality for *Phoenix* in Turkey: Gölköy-Bödrum. Principes 39, 117–122.

Bretzke, K., Armitage, S.J., Parker, A.G., Walkington, H., Uerpmann, H.P., 2013. The environmental context of Paleolithic settlement at Jebel Faya, Emirate Sharjah, UAE. Quat. Int. 300, 83–93. https://doi.org/10.1016/j.quaint.2013.01.028

Burgarella, C., Barnaud, A., Kane, N.A., Jankowski, F., Scarcelli, N., Billot, C., Vigouroux, Y., Berthouly-Salazar, C., 2019. Adaptive introgression: An untapped evolutionary mechanism for crop adaptation. Front. Plant Sci. 10, 1–17. https://doi.org/10.3389/fpls.2019.00004

Caton-Thompson, G., Gardner, E.W., 1932. The prehistoric geography of Kharga Oasis. Geogr. J. 80, 369–406. https://doi.org/10.1057/978-1-349-95943-3_742

Chao, C.T., Krueger, R.R., 2007. The Date Palm (*Phoenix dactylifera* L.): Overview of Biology, Uses and Cultivation. HortScience 42, 1077–1082.

Charbonnier, J., 2013. La maîtrise du temps d'irrigation au sein des oasis alimentées par des aflâj. Rev. d'ethnoécologie. https://doi.org/10.4000/ethnoecologie.1471

Cherif, E., Zehdi, S., Castillo, K., Chabrillange, N., Abdoulkader, S., Pintaud, J.-C.C., Santoni, S., Salhi-Hannachi, A., Glémin, S., Aberlenc-Bertossi, F., 2013. Male-specific DNA markers provide genetic evidence of an XY chromosome system, a recombination arrest and allow the tracing of paternal lineages in date palm. New Phytol. 197, 409–15. https://doi.org/10.1111/nph.12069

Chevalier, A., 1952. Les palmiers du littoral atlantique du Sud du Maroc et les Faux-Dattiers des palmeraies de Marrakech, de Tiznit et du Sous. Comptes Rendus L Acad. Des Sci. l'Académie des Sci. tome 234, 171–173.

Chevrollier, F., 2016. From Cyrene to Gortyn. Notes on the Relationships between Crete and Cyrenaica under Roman Domination (1st c. BC to 4th c. AD), in: Francis, J.E., Kouremenos, A. (Eds.), Roman Crete: New Perspectives. Oxbow Books, pp. 11–26.

Childe, V.G., 1936. Man Makes Himself, Watts & Co. ed. London.

Choi, J.Y., Platts, A.E., Fuller, D.Q., Hsing, Y.-I.I., Wing, R.A., Purugganan, M.D., Kim, Y., 2017. The rice paradox: Multiple origins but single domestication in Asian Rice. Mol. Biol. Evol. 34, 969–979. https://doi.org/10.1093/molbev/msx049

Darwin, C., 1859. On the origins of species by means of natural selection. London: Murray.

Dowson, V.H.W., Aten, A., 1962. Dates, handling, processing and packing. Rome : Food and Agriculture Organization of the United Nations.

Flowers, J.M., Hazzouri, K.M., Gros-Balthazard, M., Mo, Z., Koutrumpa, K., Perrakis, A., Ferrand, S., Khierralah, H.S.M., Fuller, D.Q., Aberlenc, F., Fournaraki, C., Purugganan, M.D., 2019. Cross-species hybridization and the origin of North African date palms. Proc. Natl. Acad. Sci. U. S. A. 116, 1651–1658. https://doi.org/10.1073/pnas.1817453116

Food and Agriculture Organization, 2013. Perennial Crops for Food Security: Proceedings of the FAO Expert Workshop; 28-30 August, 2013, Rome, Italy, in: Batelo, C., Wade, L., Cox, S., Pogna, N., Bozzini, A., Choptiany, J. (Eds.), Biodiversity & Ecosystem Services in Agicultural Production Systems. Food and Agriculture Organization, Rome, p. 390.

Fuller, D., Pelling, R., 2018. Plant Economy: Archaeobotanical Studies. Leiden, The Netherlands, pp. 349–368. https://doi.org/10.1163/9789004371583_020

Fuller, D.Q., 2018. Long and attenuated: comparative trends in the domestication of tree fruits. Veg. Hist. Archaeobot. 27, 165–176. https://doi.org/10.1007/s00334-017-0659-2

Fuller, D.Q., Stevens, C.J., 2019. Between domestication and civilization: the role of agriculture and arboriculture in the emergence of the first urban societies. Veg. Hist. Archaeobot. https://doi.org/10.1007/s00334-019-00727-4

Gardner, E.W., 1935. The Pleistocene fauna and flora of Kharga Oasis, Egypt. Q. J. Geol. Soc. 91, 479–518.

Gepts, P., 2004. Plant domestication as a long-term selection experiment. Plant Breed. Rev. 24, 1–44.

Goor, A., 1967. The history of the date through the ages in the Holy Land. Econ. Bot. 21, 320–

340. https://doi.org/10.1007/BF02863157

Gros-Balthazard, M., 2013. Hybridization in the genus *Phoenix*: A review. Emirates J. Food Agric. 25, 831–842.

Gros-Balthazard, M., Baker, W.J., Leitch, I.J., Pellicer, J., Powell, R.F., Bellot, S., 2020. Systematics and evolution of the genus Phoenix: towards understanding the date palm's origins, Preprint. https://doi.org/10.13140/RG.2.2.14666.59847

Gros-Balthazard, M., Galimberti, M., Kousathanas, A., Newton, C., Ivorra, S., Paradis, L., Vigouroux, Y., Carter, R., Tengberg, M., Battesti, V., Santoni, S., Falquet, L., Pintaud, J.-C.C., Terral, J.-F.F., Wegmann, D., 2017. The Discovery of Wild Date Palms in Oman Reveals a Complex Domestication History Involving Centers in the Middle East and Africa. Curr. Biol. 27, 2211–2218. https://doi.org/10.1016/j.cub.2017.06.045

Gros-Balthazard, M., Hazzouri, K.M., Flowers, J.M., 2018. Genomic Insights into Date Palm Origins. Genes (Basel). 9, 1–14. https://doi.org/10.3390/genes9100502

Gros-Balthazard, M., Newton, C., Ivorra, S., Pierre, M.-H.H., Pintaud, J.-C.C., Terral, J.-F.F., 2016. The Domestication Syndrome in *Phoenix dactylifera* Seeds: Toward the Identification of Wild Date Palm Populations. PLoS One 11, e0152394. https://doi.org/10.1371/journal.pone.0152394

Gros-Balthazard, M., Battesti, V., Ivorra, S., Paradis, L., Aberlenc, F., Zango, O., Zehdi, S., Moussouni, S., Naqvi, S.A., Newton, C., Terral, J., 2020. On the necessity of combining ethnobotany and genetics to assess agrobiodiversity and its evolution in crops: a case study on date palms (*Phoenix dactylifera* L.) in Siwa Oasis, Egypt. Evol. Appl. 1–23. https://doi.org/10.1111/eva.12930

Groucutt, H.S., White, T.S., Clark-Balzan, L., Parton, A., Crassard, R., Shipton, C., Jennings, R.P., Parker, A.G., Breeze, P.S., Scerri, E.M.L., Alsharekh, A., Petraglia, M.D., 2015. Human occupation of the Arabian Empty Quarter during MIS 5: Evidence from Mundafan Al-Buhayrah, Saudi Arabia. Quat. Sci. Rev. 119, 116–135. https://doi.org/10.1016/j.quascirev.2015.04.020

Harlan, J.R., 1992. Crops & Man, Foundation for Modern Crop Science Series. American Society of Agronomy.

Hazzouri, K.M., Flowers, J.M., Visser, H.J., Khierallah, H.S.M.M., Rosas, U., Pham, G.M., Meyer, R.S., Johansen, C.K., Fresquez, Z.A., Masmoudi, K., Haider, N., El Kadri, N., Idaghdour, Y., Malek, J.A., Thirkhill, D., Markhand, G.S., Krueger, R.R., Zaid, A., Purugganan, M.D., 2015. Whole genome re-sequencing of date palms yields insights into diversification of a fruit tree crop. Nat. Commun. 6, 8824. https://doi.org/10.1038/ncomms9824

Hazzouri, K.M., Gros-Balthazard, M., Flowers, J.M., Copetti, D., Lemansour, A., Lebrun, M., Masmoudi, K., Ferrand, S., Dhar, M.I., Fresquez, Z.A., Rosas, U., Zhang, J., Talag, J., Lee, S., Kudrna, D.D., Powell, R.F., Leitch, I.J., Krueger, R.R., Wing, R.A., Amiri, K.M.A.A., Purugganan, M.D., 2019. Genome-wide association mapping of date palm fruit traits.

Nat. Commun. 10, 4680. https://doi.org/10.1038/s41467-019-12604-9

Henry, A.G., Brooks, A.S., Piperno, D.R., 2011. Microfossils in calculus demonstrate consumption of plants and cooked foods in Neanderthal diets (Shanidar III, Iraq; Spy I and II, Belgium). Proc. Natl. Acad. Sci. U. S. A. 108, 486–491. https://doi.org/10.1073/pnas.1016868108

Henry, D.O., Hietala, H.J., Rosen, A.M., Demidenko, Y.E., Usik, V.I., Armagan, T.L., 2004. Human Behavioral Organization in the Middle Paleolithic: Were Neanderthals Different? Am. Anthropol. 106, 17–31. https://doi.org/10.1525/aa.2004.106.1.17

Janick, J., 2005. The origin of fruits, fruit growing and fruit breading. Plant Breed. Rev. 25, 255–320.

Kislev, M.E., Hartmann, A., Galili, E., 2004. Archaeobotanical and archaeoentomological evidence from a well at Atlit-Yam indicates colder, more humid climate on the Israeli coast during the PPNC period. J. Archaeol. Sci. 31, 1301–1310. https://doi.org/https://doi.org/10.1016/j.jas.2004.02.010

Landsberger, B., 1967. The date palm and its by-products according to the cuneiform sources.

Liphschitz, N., Nadel, D., 1997. Epipalaeolithic (19,000 B.P.) charred wood remains from Ohalo II, Sea of Galilee, Israel. Mitekufat Haeven, J. Isr. Prehist. Soc. 27, 5–18.

Malek, J.A., Mathew, S., Mathew, L., Younuskunju, S., Mohamoud, Y.A., Suhre, K., 2020. Deletion of beta-fructofuranosidase (invertase) genes is associated with sucrose content in Date Palm fruit. Plant Direct 4, 1:7. https://doi.org/10.1002/pld3.214

Malleson, C., 2016. Informal intercropping of legumes with cereals? A re-assessment of clover abundance in ancient Egyptian cereal processing by-product assemblages: archaeobotanical investigations at Khentkawes town, Giza (2300–2100 BC). Veg. Hist. Archaeobot. v. 25, 431-442–2016 v.25 no.5. https://doi.org/10.1007/s00334-016-0559-x

Malleson, C., Miracle, R., 2018. Giza Botanical Database [WWW Document]. https://doi.org/10.6078/M7JH3J99

Mathew, L.S., Seidel, M.A., George, B., Mathew, S., Spannagl, M., Haberer, G., Torres, M.F., Al-Dous, E.K., Al-Azwani, E.K., Diboun, I., Krueger, R.R., Mayer, K.F.X., Mohamoud, Y.A., Suhre, K., Malek, J.A., 2015. A Genome-Wide Survey of Date Palm Cultivars Supports Two Major Subpopulations in *Phoenix dactylifera*. G3 5, 1429–1438. https://doi.org/10.1534/g3.115.018341

McKey, D., Elias, M., Pujol, B., Duputie, A., Duputié, A., Duputie, A., 2010. The evolutionary ecology of clonally propagated domesticated plants. New Phytol. 186, 318–332. https://doi.org/10.1111/j.1469-8137.2010.03210.x

Méry, S., Tengberg, M., 2009. Food for eternity? The analysis of a date offering from a 3rd millennium BC grave at Hili N, Abu Dhabi (United Arab Emirates). J. Archaeol. Sci. 36, 2012–2017. https://doi.org/10.1016/j.jas.2009.05.017

Meyer, R.S., DuVal, A.E., Jensen, H.R., 2012. Patterns and processes in crop domestication: an historical review and quantitative analysis of 203 global food crops. New Phytol. 196, 29–48. https://doi.org/10.1111/j.1469-8137.2012.04253.x

Migicovsky, Z., Myles, S., 2017. Exploiting Wild Relatives for Genomics-assisted Breeding of Perennial Crops. Front. Plant Sci. 8, 460. https://doi.org/10.3389/fpls.2017.00460

Miller, A.J., Gross, B.L., 2011. From forest to field: perennial fruit crop domestication. Am. J. Bot. 98, 1389–1414. https://doi.org/10.3732/ajb.1000522

Mohamoud, Y.A., Mathew, L.S., Torres, M.F., Younuskunju, S., Krueger, R., Suhre, K., Malek, J.A., 2019. Novel subpopulations in date palm (*Phoenix dactylifera*) identified by population-wide organellar genome sequencing. BMC Genomics 20, 1–7. https://doi.org/10.1186/s12864-019-5834-7

Munier, P., 1981. Origine de la culture du palmier-dattier et sa propagation en Afrique. Notes historiques sur les principales palmeraies africaines. Fruits 36, 437-.

Munier, P., 1973. Le palmier-dattier, Paris: Maisonneuve et Larose.

Nixon, R.W., 1951. The Date Palm: "Tree of Life " in the Subtropical Deserts. Econ. Bot. 5, 274–301.

Parker, A.G., 2010. Chapter 10: Palaeoenvironmental evidence from H3, Kuwait, in: Carter, R.A., Crawford, H.E.W. (Eds.), Maritime Interactions in the Arabian Neolithic: The Evidence from H3, As-Sabiyah, an Ubaid-Related Site in Kuwait. Brill, Boston.

Pelling, R., 2005. Garamantian agriculture and its significance in a wider North African context: The evidence of the plant remains from the Fazzan project. J. North African Stud. 10, 397–412. https://doi.org/10.1080/13629380500336763

Peyron, G., 2000. Cultiver le palmier-dattier, Groupe de Recherche et d'Information pour le Développement de l'Agriculture d'Oasis.

Pickersgill, B., 2007. Domestication of plants in the americas: Insights from mendelian and molecular genetics. Ann. Bot. 100, 925–940. https://doi.org/10.1093/aob/mcm193

Pintaud, J.-C., Ludena, B., Zehdi, S., Gros-Balthazard, M., Ivorra, S., Terral, J.-F., Newton, C., Tengberg, M., Santoni, S., Boughedoura, N., 2013. Biogeography of the date palm (*Phoenix dactylifera* L., Arecaceae): insights on the origin and on the structure of modern diversity. ISHS Acta Hortic. 994, 19–36.

Pintaud, J.-C., Zehdi, S., Couvreur, T.L.P., Barrow, S., Henderson, S., Aberlenc-Berossi, F., Tregear, J., Billote, N., Aberlenc-Bertossi, F., Tregear, J., Billotte, N., 2010. Species delimitation in the genus *Phoenix* (Arecaceae) based on SSR markers, with emphasis on the identity of the Date Palm (*Phoenix dactylifera* L.), in: Seberg, O., Petersen, G., Barfod,

A., Davis, J. (Eds.), Diversity, Phylogeny, and Evolution in the Monocotyledons. Aarhus Univ Press, Arhus, Denmark, pp. 267–286.

Popenoe, P., 1924. The Date-palm in Antiquity. Sci. Mon. 19, 313–325.

Purugganan, M.D., Fuller, D.Q., 2011. Archaeological data reveal slow rates of evolution during plant domestication. Evolution (N. Y). 65, 171–183. https://doi.org/10.1111/j.1558-5646.2010.01093.x

Ross-Ibarra, J., Morrell, P.L., Gaut, B.S., 2007. Plant domestication, a unique opportunity to identify the genetic basis of adaptation. Proc. Natl. Acad. Sci. U. S. A. 104 Suppl, 8641–8648. https://doi.org/10.1073/pnas.0700643104

Roué, M., Battesti, V., Césard, N., Simenel, R., 2015. Ethnoecology of pollination and pollinators: Knowledge and practice in three societies TT - Ethnoécologie de la pollinisation et des pollinisateurs : savoirs et pratiques dans trois sociétés. Rev. d'ethnoécologie 0–27. https://doi.org/10.4000/ethnoecologie.2229

Sallon, S., Cherif, E., Chabrillange, N., Solowey, E., Gros-Balthazard, M., Ivorra, S., Terral, J.F., Egli, M., Aberlenc, F., 2020. Origins and insights into the historic Judean date palm based on genetic analysis of germinated ancient seeds and morphometric studies. Sci. Adv. 6, 1–11. https://doi.org/10.1126/sciadv.aax0384

Sarton, G., 1934. The Artifical Fertilization of Date-Palms in the Time of Ashur-Nasir-Pal BC 885-860. Isis 21, 8–13.

Solecki, R.S., Leroi-Gourhan, A., 1961. PALAEOCLIMATOLOGY AND ARCHAEOLOGY IN THE NEAR EAST. Ann. N. Y. Acad. Sci. 95, 729–739.

Tengberg, M., 2012. Beginnings and early history of date palm garden cultivation in the Middle East. J. Arid Environ. 86, 139–147. https://doi.org/10.1016/j.jaridenv.2011.11.022

Tengberg, M., 2003. Research into the origins of date palm domestication, in: Research, T.E.C. for S.S. and (Ed.), The Date Palm: From Traditional Resource to Green Wealth. Emirates Center for Strategic Studies and Research, Abu Dhabi, pp. 51–64.

Tengberg, M., Newton, C., 2016. Origine et évolution de la phoeniciculture au Moyen-Orient et en Egypte, in: Actes Du Colloque International Histoire Des Fruits. Pratiques Des Savoirs et Savoirs En Pratiques. Éditions Omniscience, Toulouse.

Terral, J.F., Newton, C., Ivorra, S., Gros-Balthazard, M., Tito de Morais, C., Picq, S., Tengberg, M., Pintaud, J.C., 2012. Insights into the historical biogeography of the date palm (*Phoenix dactylifera* L.) using geometric morphometry of modern and ancient seeds. J. Biogeogr. 39, 929–941.

Turner-Hissong, S.D., Mabry, M.E., Beissinger, T.M., Ross-Ibarra, J., Pires, J.C., 2020. Evolutionary insights into plant breeding. Curr. Opin. Plant Biol. 54, 93–100. https://doi.org/10.1016/j.pbi.2020.03.003

Vardareli, N., Doğaroğlu, T., Doğaç, E., Taşkın, V., Göçmen Taşkın, B., 2019. Genetic characterization of tertiary relict endemic *Phoenix theophrasti* populations in Turkey and phylogenetic relations of the species with other palm species revealed by SSR markers. Plant Syst. Evol. https://doi.org/10.1007/s00606-019-01580-8

Zaid, A., Arias-Jiménez, E.J., 1999. Date palm cultivation. FAO Plant Prod. Prot. Pap.

Zango, O., Cherif, E., Chabrillange, N., Zehdi-Azouzi, S., Gros-Balthazard, M., Naqvi, S.A., Lemansour, A., Rey, H., Bakasso, Y., Aberlenc, F., 2017. Genetic diversity of Southeastern Nigerien date palms reveals a secondary structure within Western populations. Tree Genet. Genomes 13, 75. https://doi.org/10.1007/s11295-017-1150-z

Zeder, M.A., 2017. Domestication as a model system for the extended evolutionary synthesis. Interface Focus 7. https://doi.org/10.1098/rsfs.2016.0133

Zehdi-Azouzi, S., Cherif, E., Guenni, K., Abdelkrim, A. Ben, Bermil, A., Rhouma, S., Salah, M. Ben, Santoni, S., Pintaud, J.C., Aberlenc-Bertossi, F., Hannachi, A.S., 2016. Endemic insular and coastal Tunisian date palm genetic diversity. Genetica 144, 181–190. https://doi.org/10.1007/s10709-016-9888-z

Zehdi-Azouzi, S., Cherif, E., Moussouni, S., Gros-Balthazard, M., Abbas Naqvi, S., Ludeña, B., Castillo, K., Chabrillange, N., Bouguedoura, N., Bennaceur, M., Si-Dehbi, F., Abdoulkader, S., Daher, A., Terral, J.-F., Santoni, S., Ballardini, M., Mercuri, A., Ben Salah, M., Kadri, K., Othmani, A., Littardi, C., Salhi-Hannachi, A., Pintaud, J.-C., Aberlenc-Bertossi, F., 2015. Genetic structure of the date palm (*Phoenix dactylifera*) in the Old World reveals a strong differentiation between eastern and western populations. Ann. Bot. 116, 101–112. https://doi.org/10.1093/aob/mcv068

Zhang, H., Mittal, N., Leamy, L.J., Barazani, O., Song, B.H., 2017. Back into the wild—Apply untapped genetic diversity of wild relatives for crop improvement. Evol. Appl. 10, 5–24. https://doi.org/10.1111/eva.12434

Zohary, D., Hopf, M., Weiss, E., 2012. Domestication of plants in the Old World (3rd Ed.). Oxford, UK Oxford Univ. Press.

Zohary, D., Spiegel-Roy, P., 1975. Beginnings of fruit growing in Old World. Science (80-. ). 187, 319–327. https://doi.org/10.1126/science.187.4174.319

# Boxes

**Box 1. Definitions of key words concerning the domestication of date palm**

**Artificial selection**: The process where either desirable genotypes (under conscious selection) or higher fitness genotypes (under unconscious selection) increase in frequency owing to human activity in an anthropogenic context.

**Center of domestication**: The geographic region where a crop was first domesticated. In perennial crops, domestication may be geographically diffuse and lacking distinct centers.

**Cultivation**: The activities leading to the production of food or other services from plants. The cultivation of date palm, phoeniciculture, involves only limited cultivation practices (see section 4.1).

**Domestication**: The process by which a wild crop is modified genetically and phenotypically by human activities that impact its life cycle (reproduction, propagation, selection).

**Domestication syndrome**: The phenotypic traits in a domesticated species that distinguish it from a wild species. Individual genotypes may have a mix of wild traits and domesticated traits.

**Feral**: Feral date palms are individuals originating from domesticated date palms but growing without human intervention during the life cycle. Feral populations may arise either when a cultivated population is abandoned, or when individuals 'escape' from cultivation by scattering seeds outside the fields in a natural environment that is conducive to population settlement and expansion. Although they look wild, they are not considered as genuinely wild (see 'wild').

**Introgression**: The transfer of genetic material from one population or species to another.

**Wild**: An adjective referring to a population or species of plant that has not been domesticated. The term is often appropriately applied to the relatives or progenitor (i.e., ancestor) of a domesticated crop, but wrongly applied to uncultivated populations without consideration if the ancestry of the population traces to a relative (or ancestor) or the domesticated crop (see 'Feral').

**Box 2. Categorizing and naming date palms in the garden**

Phoeniciculture involves mostly clonal propagation, and rarely planting of seeds. In palm gardens, date palms with high quality fruits are multiplied through vegetative propagation and commonly given a name (e.g., 'Deglet Noor' or 'Medjool'). Nevertheless, although such clonally propagated types are formally referred to as "cultivars" (informally as "varieties"), the nomenclature is frequently more complex in oasis agrosystems where there are three categories of named types (Battesti, 2013; Battesti et al., 2018; Gros-Balthazard et al., 2020):

- The true-to-type cultivars are names referring to palms that are indeed exclusively propagated through offshoots. Thus, they have the same morphology, produce the same quality dates, and have the same combination of alleles, except in case of somatic mutations.
- Ethnovarieties are groups of palms displaying the "same form" and the "same" dates according to farmers, but they group individuals having different combination of genes, because they are not exclusively clonally propagated. Although it's a very rare event, a seedling (from sexual reproduction) can be incorporated to the named type causing this lack of homogeneous genetic makeup among the individuals of a same named type, because, from a local perspective, it is the very same variety, the same "form".
- Local categories are groups of palms having a heterogeneous combination of genes and also heterogeneous morphologies, but are assigned a common name because of a shared characteristic. For instance, palms known to have arisen from a seed are a local category, and the name given to them depend on the locality: *khalt* in Maghreb, *úšik* in Siwa, etc. Male date palms are another category.

**Box 3. The domestication syndrome of date palms**

In date palm, the domestication syndrome is poorly described, owing to the only recent discovery of relictual wild populations that have hampered comparisons between domesticated and wild individuals. Pintaud et al. (2013) listed potential selected traits for this species, and studies of domestication syndrome in other fruit crops may provide further insights into the date palm domestication syndrome.

Characters related to the fruit and fruit production are expected to be the main targets of human selection. Overall, an increase in productivity is likely as well as a reduced inter-annual variability in yield. Domesticated dates are bigger than those of wild relatives and wild date palms (Barrow, 1998 Gros-Balthazard, pers. obs.). Seeds have indirectly been affected by human selection on the fruits (Alcántara et al., 2003; Bolmgren and Eriksson, 2010; Fuller, 2018). They are longer, their shape is more elongated (ratio length/width bigger) and extremities are more pointed (Fig. 3; Fuller, 2018; Gros-Balthazard et al., 2016; Terral et al., 2012). Fruit shape is expected to be more variable in domesticated date palms, although this has not been tested, but is visible in the seeds (Fig. 3; Gros-Balthazard et al., 2016) . Fruit flesh (mesocarp) is presumably thicker (Pintaud et al., 2013). The fruit should become more palatable with increase in sweetness, decreasing acids, and an increased variance in organoleptic characters related to selection on secondary metabolites. Other characters related to fruiting that could have been selected during the domestication process are a more dense and compact fruit bunch, a synchronic fruit maturation and delayed fruit abscission (Pintaud et al., 2013). The development of fruit without ovule fecundation (parthenocarpic fruit) has been described in other crops (e.g. fig or banana; Zohary & Spiegel-Roy, 1975; Kennedy, 2008). In date palm it would alleviate the burden of manual pollination. Yet, to our knowledge there are no cultivars that have been selected for this trait, and in most if any cases, fruits do not develop without pollination, and when they do, they don't reach maturity and are not palatable. According to Pintaud et al. (2013), the phenology may also have been affected by domestication with selection of plants having early or late flowering and fruiting to increase fruit season duration.

Vegetative organs could also have been affected by domestication. Pintaud et al 2013 proposed a reduction in basal and aerial ramification. Reduced defense of the reproductive organs through reduction of acanthophylls (spines at the base of the leaves) size that would have facilitated harvest is also a potential syndrome, as observed in the peach palm (*Bactris gasipaes*; Balick, 1984).

# Figures

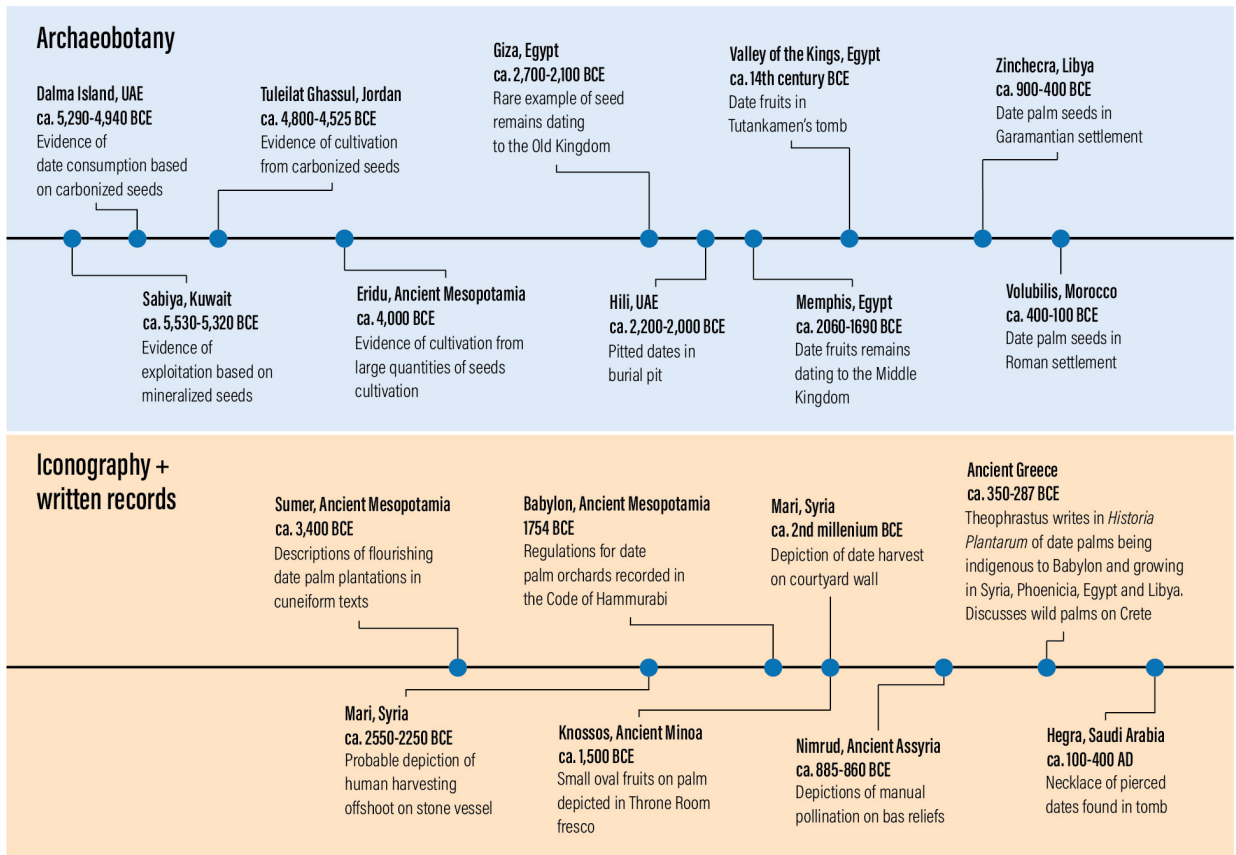

**Figure 1. A timeline of important events in the domestication history of date palms.**

Dates correspond to those reported in the references Zohary and Spiegel-Roy (1975), Tengberg et al. (2012), Malleson (2016), Malleson and Miracle (2018), Murray 1990, Mery and Tengberg (2009), Terral et al. (2012), Fuller and Pelling (2018), Aruz (2003), Zohary et al. (2012), Sarton (1934), Galanakis et al. 2017, Flowers et al. (2019).

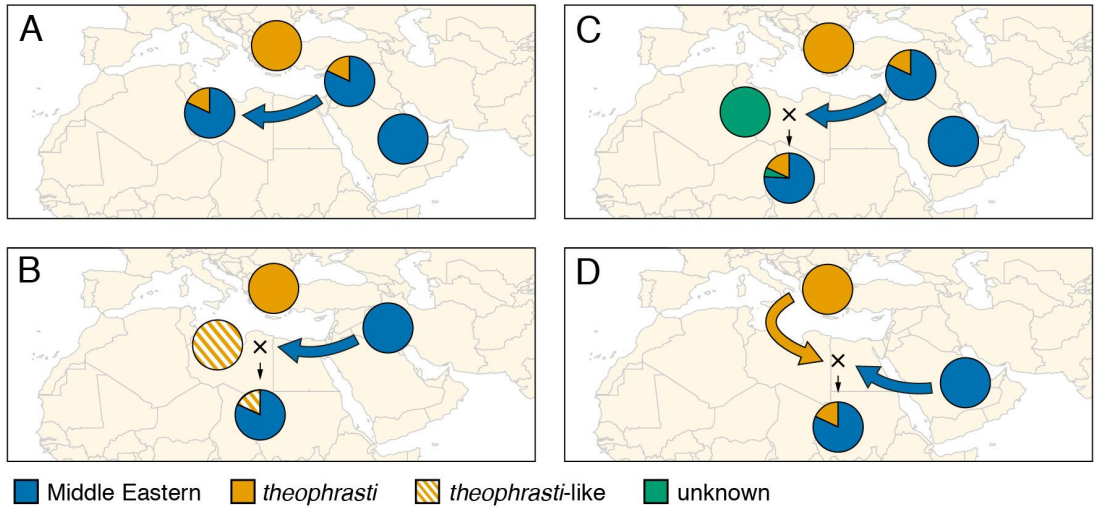

**Figure 2. Models of date palm population history.**

(A) A model where Middle Eastern date palms and *P. theophrasti* hybridized in an unknown location in the Eastern Mediterranean followed by expansion to Africa from the hybrid source. (B) A model where a resident African *P. theophrasti*-like population hybridized with Middle Eastern date palm following range expansion. Introgressive hybridization between *P. dactylifera* and resident populations of *P. theophrasti* in Greece and Turkey may not have occurred. (C) A model similar to (A) but including hybridization with a third unknown source of variation. (D) A model that invokes both the expansion of Middle Eastern date palms and transport of *P. theophrasti* to North Africa where hybridization occurred. Pie charts illustrate the genomic ancestry of a population. Thick arrows represent migration. "X" indicates introgressive hybridization. Thin arrows point to the product of introgressive hybridization.

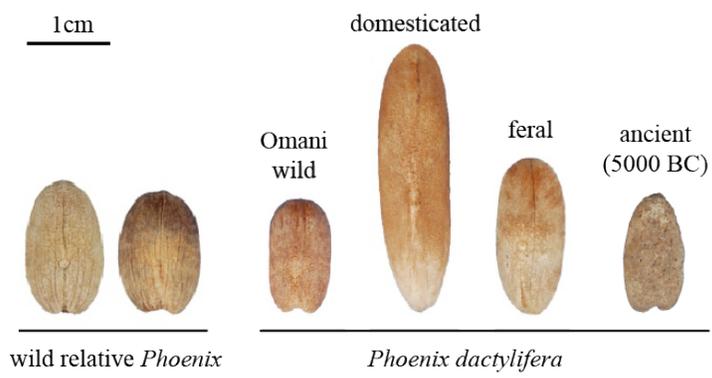

**Figure 3. Variation in seed size and shape in *Phoenix***